\documentclass{svjour3}                     
\smartqed  
\usepackage{graphicx}
\journalname{Few-Body Systems (FB20)}
\begin{document}

\title{
Few-Body Systems Composed of Heavy Quarks
\thanks{Plenary talk presented at the 20th International IUPAP Conference on Few-Body Problems
in Physics, 20 - 25 August, 2012, Fukuoka, Japan}
}


\author{Ning Li  \and  Zhi-Feng Sun \and Jun He \and Xiang Liu
\and Zhi-Gang Luo \and Shi-Lin Zhu}


\institute{Ning Li \at
              Department of Physics
and State Key Laboratory of Nuclear Physics and Technology\\
Peking University, Beijing 100871, China \\
           \and
          Zhi-Feng Sun \at
              School of Physical Science and Technology, Lanzhou
University, Lanzhou 730000,  China\\
\and Jun He \at  Nuclear Theory Group, Institute of Modern Physics
of CAS, Lanzhou 730000, China\\
\and Xiang Liu \at School of Physical Science and Technology,
Lanzhou University, Lanzhou 730000,  China\\
\and Zhi-Gang Luo\at Department of Physics
and State Key Laboratory of Nuclear Physics and Technology\\
Peking University, Beijing 100871, China\\
\and Shi-Lin Zhu \at Department of Physics
and State Key Laboratory of Nuclear Physics and Technology\\
Peking University, Beijing 100871, China\\
\email{zhusl@pku.edu.cn} }

\date{Received: date / Accepted: date}

\maketitle

\begin{abstract}

Within the past ten years many new hadrons states were observed
experimentally, some of which do not fit into the conventional
quark model. I will talk about the few-body systems composed of
heavy quarks, including the charmonium-like states and some
loosely bound states.

\end{abstract}

\section{Introduction} \label{Introduction}

QCD is the underlying theory of strong interaction, which has
three fundamental properties: asymptotic freedom, confinement, and
chiral symmetry and its spontaneous breaking. Perturbative QCD has
been tested to very high accuracy. However, the low energy sector
of QCD (i.e., hadron physics) remains very challenging.
Precision-test of Standard Model and search for new physics
requires good knowledge of hadrons as inputs (such as parton
distribution functions).

The motion and interaction of hadrons differ from those of nuclei
and quark/gluon/leptons. Hadron physics is the bridge between
nuclear physics and particle physics. The famous Higgs mechanism
contributes around 20 MeV to the nucleon mass through current
quark mass. Nearly all the mass of the visible matter in our
universe comes from QCD strong interaction. Study of hadron
physics explores the mechanism of confinement and chiral symmetry
breaking, and the mass origin.

Quark Model is quite successful in the  classification of hadrons
although it¡¯s not derived from QCD. Any state with quark content
other than $q\bar q$ or qqq is beyond naive quark model. But quark
model can¡¯t be the whole story. QCD may allow much richer hadron
spectrum such as: glueball, hybrid meson/baryon, multiquark
states, hadron molecules etc.

Experimental search of these non-conventional states started many
years ago. Typical signatures of these non-conventional states
include: (1) exotic flavor quantum number like the $\theta^+$
pentaquark; (2) exotic $J^{PC}$ quantum number like the $1^{-+}$
exotic meson; (3) overpopulation of the quark model spectrum like
the scalar isoscalar spectrum below 1.9 GeV: $\sigma$, $f_0(980)$,
$f_0(1370)$, $f_0(1500)$, $f_0(1710)$, $f_0(1790)$, $f_0(1810)$.

But till 2002, none of the above exotic candidates has been
established without controversy! Since 2003, many charmonium (or
charmonium-like) states and some Upsilon (or Upsilon-like) states
were discovered experimentally.

There are several production mechanisms of these charmonium or
charmonium-like states. First we have the initial state radiation,
where one virtual photon transforms into a $1^{--}$
charmonium(-like) state. We have the double charmonium production
where a pair of charmonium(-like) states are produced. We also
habe the B decay process. The two-photon fusion process produces
the charmonium state with even charge conjugation parity and $J\ne
1$ due to Landau-Yang theorem.

Charmonium spectrum is known very well up to the $D\bar D$
threshold. In the past decade many new states above the $D\bar D$
threshold were discovered. Some are very narrow. Some are even
charged charmonium-like states. These state are good candidates of
exotic mesons.

Many new charmonium(-like) states do not fit into quark model
spectrum easily. Theoretical speculations include: molecular
states, tetraquarks, hybrid charmonium and conventional charmonium
etc. Molecular states are loosely bound states composed of a pair
of mesons. They are probably bound by the long-range color-singlet
pion exchange. Tetraquarks are composed of four quarks. They are
bound by colored-force between quarks. They decay through
rearrangement. Some are charged. Some carry strangeness. There are
many states within the same multiplet. Hybrid charmonium are bound
states composed of a pair of quarks and one excited gluon. The
conventional quark model charmonium spectrum could also be
distorted by the coupled-channel effects.

Since some of the above states are very close to the
open-charm/bottom threshold, a quite natural interpretation is
that they could be the hadronic molecules. In QED we have the
hydrogen atom. One light electron circles around the proton. When
two electrons are shared by two protons, a hydrogen molecule is
formed. In QCD we have the heavy meson and baryon. One light quark
circles around the heavy anti-quark. Two light quarks circle
around the heavy quark. Do we expect the analogue of the hydrogen
molecule in QCD? I will focus on the hadron molecules composed of
a pair of heavy mesons or heavy baryons in this talk.

The idea of the loosely bound molecular states is not new in
nuclear physics since Yukawa proposed the pion in 1935. The
deuteron is a very loosely bound state composed of a proton and
neutron arising from the color-singlet meson exchange. Besides the
long-range pion exchange, the medium-range attraction from the
correlated two-pion exchange (or in the form of the sigma meson
exchange), the short-range interaction in terms of the vector
meson exchange, and the S-D wave mixing combine to form the
loosely bound deuteron. We adopt the same one-boson-exchange (OBE)
formalism to discuss the possible molecular states composed of a
pair of heavy mesons or heavy baryons.

The interaction potential is derived assuming the hadrons are
point-like particles. For a loosely bound state, the pion does NOT
explore the inner structure of the hadron. In other words, the
momentum of the pion is soft. Therefore, a form factor is always
introduced at each vertex to regulate the potential and suppress
the contribution from the ultraviolet regime of the momentum. The
cutoff in the form factor is constrained by the deuteron data.

Before we start, we should keep one lesson in mind. In 1949, Fermi
and Yang derived the $N\bar N$ potential based on $N\bar N\to
N\bar N$ elastic scattering and G-parity rule. They obtained many
deeply bound states in the OBE model, but none of them was
observed experimentally. Why? In the $N\bar N$ case, the inelastic
scattering or annihilation process $N\bar N \to \mbox{mesons}$ is
also important, which renders the short-distance interaction very
complicated. In fact, the presence of the optical potential V(r)+i
W(r) changes the whole picture. Generally speaking, the OBE model
is reliable only in the case (1) when there is no annihilation;
(2) or for very shallow bound states when there is annihilation.

\section{The charged $Z_b$ states}\label{Zb}

Now let's move on to the charged $Z_b$ states. In 2011, the Belle
Collaboration announced two charged Upsilon-like states
$Z_b(10610)$ and $Z_b(10650)$. These two states were observed in
the invariant mass spectra of $h_b(nP)\pi^\pm$ ($n=1,2$) and
$\Upsilon(mS)\pi^\pm$ ($m=1,2,3$) of the corresponding
$\Upsilon(5S)\to h_b(nP)\pi^+\pi^-$ and $\Upsilon(5S)\to
\Upsilon(mS)\pi^+\pi^-$ hidden-bottom decays
\cite{Collaboration:2011gj}. With the above five hidden-bottom
decay channels, Belle extracted the $Z_b(10610)$ and $Z_b(10650)$
parameters. The obtained averages over all five channels are
$M_{Z_b(10610)}=10608.4\pm2.0$ MeV/c$^2$,
$\Gamma_{Z_b(10610)}=15.6\pm2.5$ MeV/c$^2$,
$M_{Z_b(10650)}=10653.2\pm1.5$ MeV/c$^2$,
$\Gamma_{Z_b(10650)}=14.4\pm3.2$ MeV/c$^2$
\cite{Collaboration:2011gj}. In addition, the analysis of the
angular distribution indicates both $Z_b(10610)$ and $Z_b(10650)$
favor $I^G(J^P)=1^+(1^+)$.

If $Z_b(10610)$ and $Z_b(10650)$ arise from the resonance
structures, they are good candidates of non-conventional
bottomonium-like states. The masses of the $J^{PC}=1^{++}$ and
$J^{PC}=1^{+-}$ $b\bar b q\bar q$ tetraquark states were found to
be around $10.1\sim 10.2$ GeV in the framework of QCD sum rule
formalism \cite{chenwei}, which are significantly lower than these
two charged $Z_b$ states. Therefore, it's hard to accommodate them
as tetraquarks. If comparing the experimental measurement with the
$B\bar{B}^*$ and $B^*\bar{B}^{*}$ thresholds, one notices that
$Z_b(10610)$ and $Z_b(10650)$ are close to thresholds of
$B\bar{B}^*$ and $B^*\bar{B}^{*}$, respectively. One plausible
explanation is that both $Z_b(10610)$ and $Z_b(10650)$ are either
$B^*\bar{B}^{*}$ or $B^*\bar{B}^{*}$ molecular states
respectively.

Before the observations of two charged $Z_b(10610)$ and
$Z_b(10650)$ states, there have been many theoretical works which
focused on the molecular systems composed of $B^{(*)}$ and
$\bar{B}^{(*)}$ meson pair and indicated that there probably exist
loosely bound S-wave $B^*\bar{B}^{*}$ or $B^*\bar{B}^{*}$
molecular states \cite{Liu:2008fh,Liu:2008tn}. To some extent,
such studies were stimulated by a series of near-threshold
charomonium-like $X$, $Y$, $Z$ states in the past eight years.

Within the OBE framework, one can construct the effective
potentials arising from various meson exchange contributions. We
need also consider the S-wave and D-wave mixing contribution.
After solving the couple-channel Schrodinger equation, we found
that both $Z_b$(10610) and $Z_b$(10650) can be explained as the
BB* and B*B* molecular states.  Besides the isovector $Z_b$
states, there are also several loosely bound isoscalar molecular
states \cite{LIU_Zb}.

Besides the five hidden-bottom discovery modes, Belle
collaboration confirmed these two charged states in the $B(*){\bar
B}*$ channel in July 2012 \cite{Belle-July}. In fact these are the
dominant decay modes. The $B{\bar B}*$ branching ratio is
$(86.0\pm 3.6)\%$ for the lower $Z_b$ state. For the heavy $Z_b$
state, the $B(*){\bar B}*$ branching ratio is $(73.4\pm 7.0)\%$.
Moreover, Belle collaboration reported the first evidence of the
neutral partner of these $Z_b$ states \cite{Belle-neutral}. Up to
now, the properties of the $Z_b$ states fit the molecular
hypothesis quite well.

The masses of the final states $B{\bar B}*\pi$ and $B*{\bar
B}*\pi$ are 10.744 GeV and 10.790 GeV respectively, which are very
close to the $\Upsilon (5S)$ mass 10.860 GeV. In other words, the
decay phase space is tiny. The relative motion between the B(*)B*
pair is very slow, which is favorable to the formation of the
B(*)B* molecular states. To some extent, $\Upsilon (5S)$ [or
$\Upsilon (6S)$] is the ideal factory of the heavy molecular
states, which will be produced abundantly at sBelle in the near
future!

\section{X(3872)}\label{X3872}

The narrow state X(3872) was also discovered by Belle
collaboration. Its width is extremely narrow, i.e., less than 1.2
MeV. Its mass is very close to the $D*\bar D$ threshold. No
charged partners were found up to now. So it's not an isovector.
This state decays into $J/\psi \gamma$. Hence its C-parity is
even. Both CDF and Belle collaborations performed the angular
distribution analysis and found its quantum number could be
$1^{++}$ or $2^{-+}$. Both Belle and Babar collaborations observed
large $D^0 {\bar D}^{\ast 0}$ branching ratio. The most puzzling
issue of X(3872) is the large isospin symmetry breaking in its
hidden-charm decays. The dipion hidden-charm decay violates
isospin symmetry, but its decay width is comparable to that of the
three-pion decay mode, which conserves isospin symmetry. As a
$2^{-+}$ or $1^{++}$ charmonium state, it is very difficult to
explain the large isospin symmetry breaking around 100\%. The
typical isospin symmetry breaking effect from up/down quark mass
difference and QED is around $0.1\%\to 1\%$.

In fact, the proximity of the $X(3872)$ to the threshold of
$D^0\bar{D}^{*0}$ strongly suggests that the $X(3872)$ is probably
a loosely bound $D^0\bar{D}^{*0}$ molecule. We extend the OBE
model to study X(3872) as a $1^{++}$ molecular state.

We want to find out the specific role of the charged $D\bar{D}^*$
mode, the isospin breaking and the coupling of the $X(3872)$ to
$D^*\bar{D}^*$ in forming the loosely bound $X(3872)$. We first
consider the neutral component $D^0\bar{D}^{*0}$ only and include
the S-D wave mixing, which corresponds to Case (I). Then we add
the charged $D^+D^{*-}$ component to form the exact $D\bar{D}^*$
isospin singlet with the S-D mixing, which is Case (II). Since the
$1^{++}$ $D^*\bar{D}^*$ channel lies only 140 MeV above and
couples strongly to the $D\bar{D}^*$ channel, we further introduce
the coupling of $D\bar{D}^*$ to $D^*\bar{D}^*$ in Case (III).
Finally, we move one step further and take into account the
explicit mass splitting between the charged and neutral
$D(D^\ast)$ mesons, which is the physical Case (IV).

In case (IV) we consider the isospin breaking for $D\bar{D}^*$
only but keep the isospin limit for the $D^*\bar{D}^*$ channel.
Since the threshold of $D^*\bar{D}^*$ is about 140 MeV above the
$X(3872)$ mass, the probability of the $D^*\bar{D}^*$ component is
already quite small due to such a large mass gap. The isospin
breaking effect due to the mass splitting of the $D^*\bar{D}^*$
pair is even smaller and negligible. In case (IV) we have omitted
the channel ${1\over
\sqrt{2}}\left(D^{*0}\bar{D}^{*0}+D^{*+}D^{*-}\right)|^5D_1>$. At
the first glimpse, this channel should also be included. After
careful calculation, it turns out that the matrix elements between
this channel and other channels are zero.

Let's look at the results in one-pion-exchange (OPE) model first.
Now there is only one coupling constant g. Its value g=0.59 was
extracted from the decay width of the D* meson. In Case I, we find
no binding solutions with the cutoff parameter around 0.8~2.0 GeV.

After adding the charged mode and assuming they are degenerate
with the neutral mode, we obtain a loosely bound state with
binding energy 0.32 MeV for the cutoff parameter being 1.55 GeV in
Case II. The root-mean-square radius is 4.97 fm The S wave is
dominant, with a probability of 98.81\% while
the contribution of the D wave is 1.19%
The inclusion of the charged mode enhances the attraction by a
factor of three. The charged mode is important in the formation of
the bound state, although the required cutoff parameter is larger
than 1.5 GeV.

After we include the coupling of DD* to D*D*, we can see the
important role of the coupled-channel effects in Case III. The
binding energy increases by several tens MeV compared with Case II
with the same cutoff parameter. The binding energy is 0.76 MeV and
the root-mean-square radius is 3.79 fm with the cutoff parameter
around 1.10 GeV, which is a reasonable value.

Does the state in Case I-III correspond to X(3872)?  The answer is
No. The state in Case I only contains the neutral component, which
is an equal superposition state of the isoscalar and isovector
state. The states in Cases II and III are definitely isoscalar.
Experimentally, the hidden-charm di-pion decay mode of X(3872)
violates isospin symmetry. Actually, none of the above states
corresponds to the physical state of X(3872).

We further consider the mass splitting of the neutral and charged
D(D*) mesons in Case IV. Now the binding energy decreases by
roughly 3 MeV compared to Case III with the same cutoff parameter.
This is reasonable since the charged DD* pair is almost 8 MeV
heavier than the neutral pair. Now the flavor wave function of
this very loosely bound molecular state contains a large isovector
component, which decays into the $J/\psi \rho$ mode. This
molecular state can be interpreted as X(3872).

We may also take into account the heavier meson exchange as well
as the pion exchange. When the cutoff parameter is fixed at 1.05
GeV, the heavier meson exchange potentials cancel each other to a
large extent. Therefore, the long-range pion exchange still plays
a dominant role in forming the loosely bound state. The residual
effect of the heavier meson exchange modify the binding solution
slightly.

The contribution of the isoscalar component is 74\% while that of
the isovector component is 26\% if the binding energy is 0.3 MeV
when the cutoff parameter is fixed at 1.05 GeV. However, if the
binding energy increases to ~11 MeV, the contribution of the
isoscalar component is as large as ~98.5\% while that of the
isovector component is only 1.5\%. The isospin breaking depends
sensitively on the binding energy. The large isospin symmetry
breaking effect in the flavor wave function is amplified by the
tiny binding energy.

There is experimental evidence that the hidden-charm di-pion decay
occurs through a virtual $\rho$ meson while the hidden-charm
three-pion decay occurs through a virtual $\omega$ meson. The
isospin violating di-pion decay comes from the isovector component
within the flavor wave function while the three-pion decay comes
from the isoscalar component. The different phase space of the
$J/\psi \rho$ and $J/\psi \omega$ decay modes also plays an
important role. The ratio of the above two phase spaces is 0.15.

The branching fraction ratio
\begin{equation}
R=R_{Phase}\times R_I = \mathcal{B}(X(3872)\to
\pi^+\pi^-\pi^0J/\psi)/\mathcal{B}(X(3872)\to \pi^+\pi^-
J/\psi)=0.42
\end{equation}
with the binding energy being 0.3 MeV. Again, this ratio depends
very sensitively on the binding energy since the isospin breaking
effect is very sensitive to the binding energy. Given the
uncertainty of experimental value of the mass of $X(3872)$, this
ratio is consistent with the experimental value,
$1.0\pm0.4(stat)\pm 0.3(syst)$ from Belle Collaboration
~\cite{Abe:2005ix} and $0.8\pm 0.3$ from BABAR Collaboration
~\cite{delAmoSanchez:2010jr}.

In short summary, the existence of the loosely bound state X(3872)
and the large isospin symmetry breaking in its hidden-charm decay
arises from the very delicate efforts of the several driving
forces including: long-range one-pion exchange, S-D wave mixing,
mass splitting between the charged and neutral D(D*) mesons,
coupled-channel effects. The extreme sensitivity of the physical
observables to the tiny binding energy is typical of the loosely
bound system.

\section{Heavy Dibaryons}\label{baryon}

We extend the same OBE formalism to the heavy di-baryon system
which roughly reproduce the qualitative features of the deuteron
(B.E. ~2 MeV, radius ~ 2 fm). The ground state heavy baryons (qqQ)
include one spin-1/2 and -3/2 sextet and spin-1/2 triplet. Among
these states, $\Lambda_Q, \Xi_Q, \Xi'_Q$ are stable with weak
decays only. $Q=b$ or $c$ denotes the corresponding heavy quark.
The pion heavy baryon coupling constants are related to the
well-known pion-N-N coupling constants with the pion-quark-quark
coupling in the chiral quark model as the bridge.

We focus on the heavy H di-baryon and perform a coupled channel
analysis of the $\Lambda_Q \Lambda_Q$ system. For the scalar
isoscalar channel, one need consider the channel couplings between
$\Lambda_Q\Lambda_Q (^1S_0), \Sigma_Q \Sigma_Q (^1S_0),
\Sigma^\ast_Q\Sigma^\ast_Q (^1S_0), \Sigma_Q\Sigma^\ast_Q(^5D_0),
\Sigma^\ast_Q\Sigma^\ast_Q(^5D_0)$ \cite{lee1,oka,lee2}.

With the couple channel effects, the one pion exchange force alone
is strong enough to bind the $\Lambda_Q\Lambda_Q$ system. As the
heavy quark mass increases, the binding becomes deeper. For the
heavy analogue of the H dibaryon, our results indicate that
$\Lambda_Q\Lambda_Q$(Q=b,c) with quantum numbers $I(J^P)=0(0^+)$,
$0(0^-)$ and $0(1^-)$ may all be molecules. The binding solutions
of $\Lambda_Q\Lambda_Q$ system with the OPE potential mainly come
from the coupled-channel effect. Besides the transition induced by
the OPE force in the flavor space, we have also considered the
transitions caused by the eta meson and rho/omega meson exchange.
With the same cutoff parameter, the binding energy in the OBE case
is larger than that in the OPE case. The medium- and short-range
attractive force plays a significant role in the formation of the
loosely bound $\Lambda_c\Lambda_c$ and $\Lambda_b\Lambda_b$
states.

The heavy analogue of the H dibaryon probably exists due to the
coupled channel effects while the light H dibaryon
($\Lambda\Lambda$) is unbound or barely bound. In fact, the
long-range OPE force is strong enough to form the loosely bound
$\Lambda_Q\Lambda_Q$ states. As the heavy quark mass increases,
the kinetic energy of the system decreases while the potential
remains roughly the same, which is favorable to the formation of
the shallow bound states. Once produced, the heavy H dibaryons may
decay into the $N\Xi_{cc}$ final state \cite{jmr}. Such a
transition requires the exchange of a heavy D/B meson between the
$\Lambda_Q$ pair, which occurs at very short distance. On the
other hand, the $\Lambda_Q$ pair is loosely bound with a large
radius. In other words, the decay width of this decay mode is
expected to be small. Maybe these states could be produced at
RHIC, LHC?

\section{Summary}

In the past decade, many charmonium(-like) and some Upsilon(-like)
states were discovered. Some states do not fit into the quark
model spectrum easily. Some of them lie very close to the
open-charm/bottom threshold with narrow width such as X(3872).
Some are even charged like the $Z_b$ states. These states are very
good candidates of loosely bound hadron molecules In the near
future, LHCb, J-PARC, sBelle, Pande will contribute to the search
of non-conventional hadrons. We expect more unexpected.

\begin{acknowledgements}
This project was supported by the National Natural Science
Foundation of China under Grants 11075004, 11021092 and Ministry
of Science and Technology of China (2009CB825200). This work is
also supported in part by the DFG and the NSFC through funds
provided to the sino-germen CRC 110 ``Symmetries and the Emergence
of Structure in QCD''.
\end{acknowledgements}


\end{document}